\documentclass[12pt]{article}
\usepackage{epsfig}
\usepackage{graphicx}
\usepackage{amsmath}
\usepackage{amssymb}
\usepackage{slashed}
\usepackage{epstopdf}
\usepackage{cite}

\usepackage{xcolor}
\definecolor{lcolor}{rgb}{0.,0.0,0.}
\definecolor{citcolor}{rgb}{0,0.,0.5}
\usepackage[breaklinks,colorlinks,urlcolor=blue,citecolor=blue,linkcolor=blue]{hyperref}
\usepackage{mciteplus}

\numberwithin{equation}{section}

\newcommand{\beq}{\begin{equation}}
\newcommand{\eeq}{\end{equation}}
\newcommand{\bea}{\begin{eqnarray}}
\newcommand{\eea}{\end{eqnarray}}
\newcommand{\dis}{\displaystyle}
\def\dd{{\rm d}}

\newcommand{\bem}{\begin{multline}}
\newcommand{\eem}{\end{multline}}
\newcommand{\beg}{\begin{gather}}
\newcommand{\eeg}{\end{gather}}

\newcommand{\nn}{\nonumber}

\newcommand{\ben}{\begin{eqnarray*}}
\newcommand{\een}{\end{eqnarray*}}

\newcommand{\eq}[1]{\begin{align}#1\end{align}}

\begin{document}
\title{{\bf The neutrino-nucleon cross section at UHE and its astrophysical implications\\[0.5cm] }}

\author{
{Javier L. Albacete, Jos\'e I. Illana, Alba Soto-Ontoso }\\[0.2cm]  {\it \small CAFPE and Departamento de F\'isica Te\'orica y del Cosmos,  Universidad de Granada}\\ {\it \small E-18071 Campus de Fuentenueva, Granada, Spain.} \\[0.1cm] {\texttt{ \small albacete@ugr.es, jillana@ugr.es, aontoso@ugr.es}}
  }
\date{\small UG-FT-316/15\qquad CAFPE-186/15}
\maketitle

\begin{abstract}
 
  We present a quantitative study of the $\nu N$ cross section in the neutrino energy range $10^4<E_{\nu}<10^{14}$~GeV within two transversal QCD approaches: NLO DGLAP evolution using different sets of PDFs and BK small-$x$ evolution with running coupling and kinematical corrections. We show that the non-linear effects embodied in the BK equation yield a slower raise in the cross section for $E_{\nu}\gtrsim 10^{8}$~GeV than the usual DGLAP based calculation. Finally, we translate this theoretical uncertainty into upper bounds for the ultra-high-energy neutrino flux for different experiments.
\end{abstract}

\section{Introduction} 

The observation of neutrino events with energies in the order of PeV by the IceCube Observatory \cite{Aartsen:2013bka} has opened a new era in Neutrino Physics.
Currently ongoing and proposed experimental programs aim at the determination of the neutrino flux for energies of the incident neutrinos several orders of magnitude higher than those measurable at IceCube. Thus, the Auger collaboration has recently reported on the limits of neutrino fluxes at ultra high-energy (UHE) in \cite{Aab:2015kma}.  Also, the LUNASKA project \cite{James:2009sf} performs related measurements using the Square Kilometre Array \cite{Carilli:2004nx}. Among the several astrophysical scenarios proposed as source of UHE neutrinos, the interaction of UHE cosmic rays with the cosmic microwave background would result in a non-negligible flux of neutrinos impinging on Earth up to energies of $E_{\nu}\sim 10^{12-14}$~GeV.

An essential ingredient for the determination of neutrino fluxes from experimental data is the precise knowledge of the neutrino-nucleon ($\nu N)$ cross section, as it relates the neutrino flux with the number of observed events at a given neutrino energy. However, the study of the $\nu N$ cross section ($\sigma_{\nu N}$) at UHE implies probing QCD in a kinematic regime unexplored so far in ground based accelerators, like HERA or the LHC. This region is characterized by very small values of Bjorken-$x$ ($x\lesssim 10^{-7}$ for $E_{\nu}\gtrsim 10^{11}$~GeV) and virtualities of the order of the electroweak boson mass squared, $Q^2\sim M_{Z,W}^2\sim10^4$~GeV$^2$ (for a more precise discussion of the kinematics, see Section~\ref{2}). Thus, a reliable calculation of $\sigma_{\nu N}$ in the previously uncharted kinematic territory relevant for UHE neutrino interactions requires theoretically controlled extrapolations from the well tested region of phase space studied in collider experiments.

Calculating the cross section of any high-energy hadronic collision, like the $\nu N$ studied in this work, requires a priori knowledge of the partonic structure of the proton at all relevant observation scales. In practice such information is provided by phenomenological parton fits to previously existing data based on the use of perturbative QCD renormalization group equations in the framework of well defined factorization theorems.

The different QCD approaches for the description of the scale dependence of parton distribution functions -- analogously, for gauge invariant operators encoding the parton flux into the collision -- share a similar strategy of resumming radiative terms enhanced by large logarithms to all orders. 
The most widely used framework are the DGLAP equations~\cite{Dokshitzer:1977sg,Gribov:1972ri,Altarelli:1977zs}, that is, the renormalization group equations that describe the scale dependence of parton distribution functions through a resummation of large logarithms $\sim \alpha_{s}\ln Q^2/Q_0^2$, with $Q_{0}$ some initial scale. 
The DGLAP equations have been successfully and intensively tested against experimental data and, together with asymptotic freedom and factorization theorems,  provide a fundamental tool for establishing controlled theoretical predictions. Successful as they are, the DGLAP equations are also expected to break down in some kinematic regimes. In particular, at small values of Bjorken-$x$, large logarithms $\sim\alpha_{s}\ln(x_{0}/x)$ emerge and also need to be resummed to all orders. 

In turn, analogous resummation schemes aimed at describing the small-$x$ evolution of hadron structure have also been developed. In this direction in the kinematic $(x,Q^2)$-plane, orthogonal to DGLAP evolution, the relevant logarithms are $\sim\alpha_{s}\ln(x_{0}/x)$, resummed to all orders in the BFKL approach~\cite{Kuraev:1977fs,Balitsky:1978ic}. 
Additionally, the enhancement of gluon emission at small-$x$ naturally leads to the presence of large gluon densities in the proton and to the need of non-linear {\it recombination} terms in order to stabilize the diffusion towards the infrared characteristic of BFKL evolution. Most importantly, the presence of non-linear terms is ultimately related to the preservation of unitarity of the theory.  
Both the resummation of small-$x$ logarithms and the inclusion of non-linear density dependent corrections are consistently accounted for by the Balitsky-Kovchegov (BK) equation \cite{Kovchegov:1999yj,Balitsky:1996ub}, which corresponds to the large-$N_c$ limit of the B-JIMWLK~\cite{Balitsky:1996ub,Jalilian-Marian:1997gr,Kovner:2000pt,Weigert:2000gi} hierarchy of coupled non-linear evolution equations. The presence of non-linear terms in the small-$x$ evolution equations limits the growth rate of gluon number densities for modes of transverse momentum smaller than the saturation scale $Q_{s}$ (see Section~\ref{bk} for a precise definition). This novel, semi-hard dynamical scale marks the onset of non-linear corrections in QCD evolution and leads to distinctive dynamical effects such as the generation of geometric scaling~\cite{Stasto:2000er}. The ability of the BK equation at running coupling accuracy to describe the Bjorken-$x$ dependence of the  structure functions in Deeply Inelastic electron-proton Scattering (DIS) measured at HERA has been well established in a series of recent works \cite{Albacete:2009fh,Albacete:2010sy,Lappi:2013zma}. Further, it was shown in \cite{Albacete:2012rx} that BK-based fits to low-$x$ HERA data are more stable than analogous DGLAP fits under the change in the boundary conditions.  The unintegrated gluon distributions resulting from these BK global fits to HERA data have found many successful phenomenological applications in the analysis of data from the proton-proton, proton-nucleus and nucleus-nucleus experimental programs at RHIC and the LHC (see e.g.\cite{Albacete:2014fwa} for a review), providing strong evidence of the presence of non-linear saturation effects in available experimental data.

The natural question arises of which of these two {\it orthogonal} QCD approaches, DGLAP or BK, is better suited to extrapolate our well stablished knowledge of parton structure to the region of phase space proben in UHE neutrino of very small-$x$ and high $Q^2$ values.
Clearly, the reliability of either approach in this intermediate kinematic region cannot be determined on a priori theoretical arguments. This is so because, at a parametric level, one expects that both large logarithmic corrections $\sim \ln Q^2$ and $\ln 1/x$ resumed in either approach to be relevant in that kinematic regime.  It is also clear that claims in favor of one particular approach should not be done solely on the basis of agreement with previous experimental data: it is well known that one can obtain an excellent fit to the HERA low-$x$ data with
a very reduced number of free parameters. Such is a necessary but not sufficient condition.  Further, beyond describing existing data, the usefulness of a given approach rests on its predictive power towards kinematic regions experimentally unexplored so far. This latter condition lessens the reliability  of phenomenological models not equipped with a well defined QCD dynamical input. In our view the use of one or another approach should be considered as a systematic uncertainty associated to the theoretical estimate to the $\nu N$ cross section. It is the goal of this work to provide a precise quantitative reference for such systematic uncertainty.  As we shall explain in detail in the following sections, we find that the differences arising in the calculation of the neutrino-nucleon cross section at UHE due to the choice of either the DGLAP or BK approaches are sizeable for energies of the incident neutrino $E_\nu \gtrsim 10^{8} $~GeV. This differences become as large as a factor 4.5 for the highest neutrino energies of $E_\nu = 10^{14}$~GeV studied here. In line with our expectations, the values of the $\nu N$ total cross sections obtained within the BK evolution approach are systematically smaller than those obtained within the DGLAP approach due to the presence of non-linear recombination effects accounted for the BK approach.

This paper is organized as follows. In the next section a brief review of the underlying formalism of neutrino deep inelastic scattering is presented. A systematic study of the DGLAP approach at NLO has been performed in Section~\ref{dglap} including the corresponding error bands both for charged and neutral current interactions. In \ref{bk} we introduce state-of-the art saturation effects in the computation of $\sigma_{\nu N}$ through numerical solutions of the BK equation including running coupling and kinematical corrections. Finally, as a phenomenological application we recalculate the limits of the UHE neutrino flux obtained by various experiments with our new parametrization for the cross section. 

\section{Neutrino-nucleon cross section}\label{2}

The inelastic interaction of neutrinos with nucleons is described in terms of charged current (CC) and neutral current (NC) interactions, which proceed through $W^{\pm}$ and $Z^0$ exchanges, respectively. The expression of the differential cross section in the fixed-target frame is \cite{devenish2004deep}:
\eq{
\dis\frac{\dd^2\sigma ^{\rm{CC,NC}}_{\nu N}}{\dd x\dd y}&=
\dis\frac{G_{F}^2 s}{\pi}
\left(\dis\frac{M_{i}^2}{M_{i}^2+Q^2}\right)^2 \nn\\&\times
\left[
xy^2F_1^{\rm{CC,NC}}(x,Q^2)+(1-y)F_2^{\rm{CC,NC}}(x,Q^2)
+ xy\left(1-\dis\frac{y}{2}\right)F_{3}^{\rm{CC,NC}}(x,Q^2)\right]
\label{dif-xs}
}
where $G_F$ is the Fermi coupling constant, $E_{\nu}$ is the neutrino energy, $s=2M_{N}E_{\nu}$ with $M_{N}$ the nucleon mass and $M_i$ denotes the mass of the charged or neutral gauge boson exchanged. The kinematics of this process is described in terms of   the virtuality of the gauge boson $Q^2$, the fraction of the nucleon's momentum carried by the struck quark Bjorken-$x$ and the inelasticity $y=Q^2/(xs)$. The extension of our results to the case of neutrino-nucleus cross section is straightforward. Nuclear effects like nuclear shadowing or other destructive interference effects can be neglected at high virtualities (see e.g. the nuclear PDF sets in \cite{Eskola:2008ca}). This allows to consider nuclei as an ensemble of $A$ independent nucleons, leading to a trivial scaling of the total cross section with the atomic number $A$. Further, we will assume an isoscalar target $N\equiv(p+n)/2$ which is a good approximation to the typical neutrino detector material. 

The integral of the differential cross section Eq. \ref{dif-xs} is dominated by values of $Q^2\sim M_i^2=10^{4}$~GeV$^2$. Hence the typical $x$ value probed is $x_{\rm{min}}\sim M_i^2/2M_NE_{\nu}$. For $E_{\nu}>10^8$~GeV this translates into $x_{\rm{min}}<10^{-5}$, that is, for ultrahigh energy neutrinos we are going to work in the region of small-$x$ and $Q^2$ values of $\sim M_{Z,W}^2$.

Once the underlying formalism has been presented, in the following sections we are going to compute the structure functions $F_i(x,Q^2)$ in both the DGLAP and BK frameworks.

\subsection{Improved parton model and DGLAP evolution}\label{dglap}

In the QCD improved parton model the structure functions $F_i(x,Q^2)$ are linear combinations of parton distribution functions. Explicit expressions for $F_i(x,Q^2)$ both at leading order (LO) and next-to-leading order (NLO) are given in Appendix \ref{nlo}. The contribution of top sea quarks is completely negligible in the phase space region of our study \cite{Gandhi:1998ri}. So it is the contribution of $F_3^{\rm{CC,NC}}$ in the energy range that we are dealing with, therefore we are not going to consider it in our further numerical calculations. We have chosen two sets of PDFs, {\tt NNPDF3.0} \cite{Ball:2014uwa} and {\tt MSTW08} \cite{Martin:2009iq}. Both sets of parton distribution function correspond to DGLAP evolution at NLO accuracy. Even though the two PDF sets rely on the same theoretical framework, differences in the implementation like, for instance, the choice of initial conditions for the evolution, and other details result in slightly different values of the parton distribution functions.

\begin{figure}[htb]
\begin{center}
\includegraphics[scale=0.8]{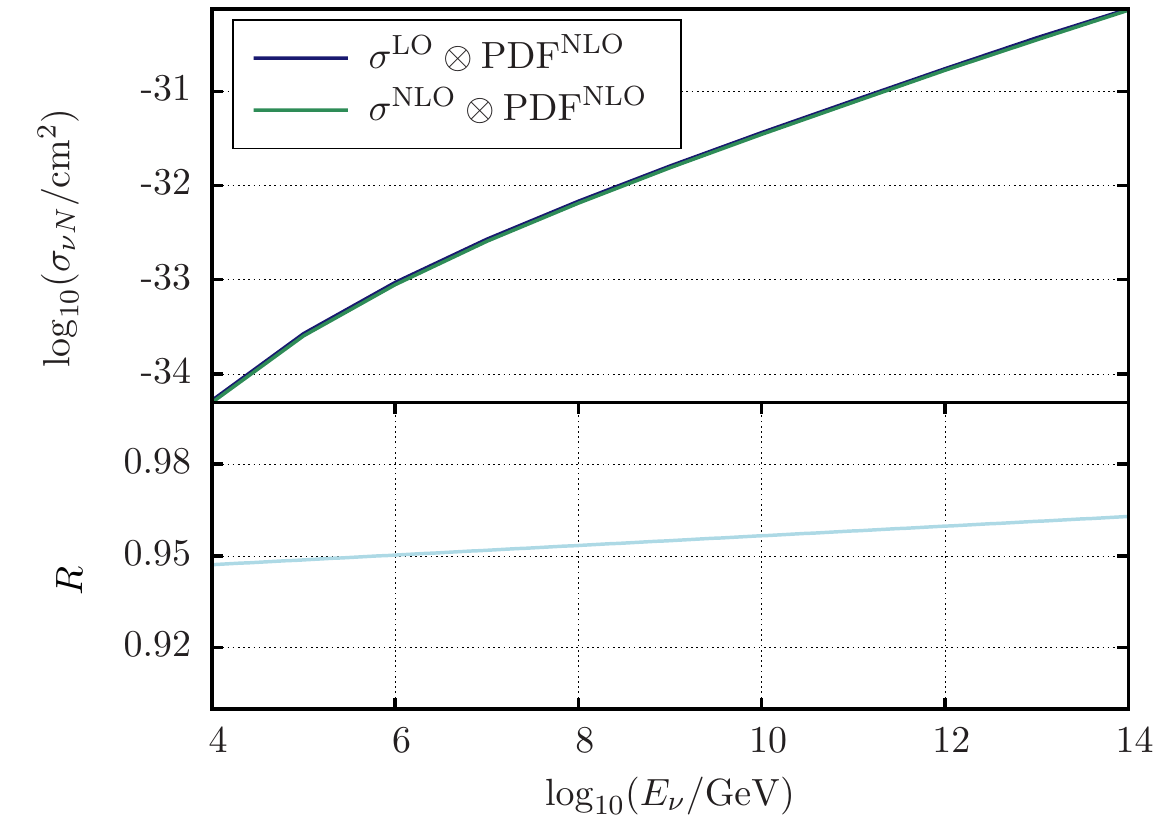}
\includegraphics[scale=0.8]{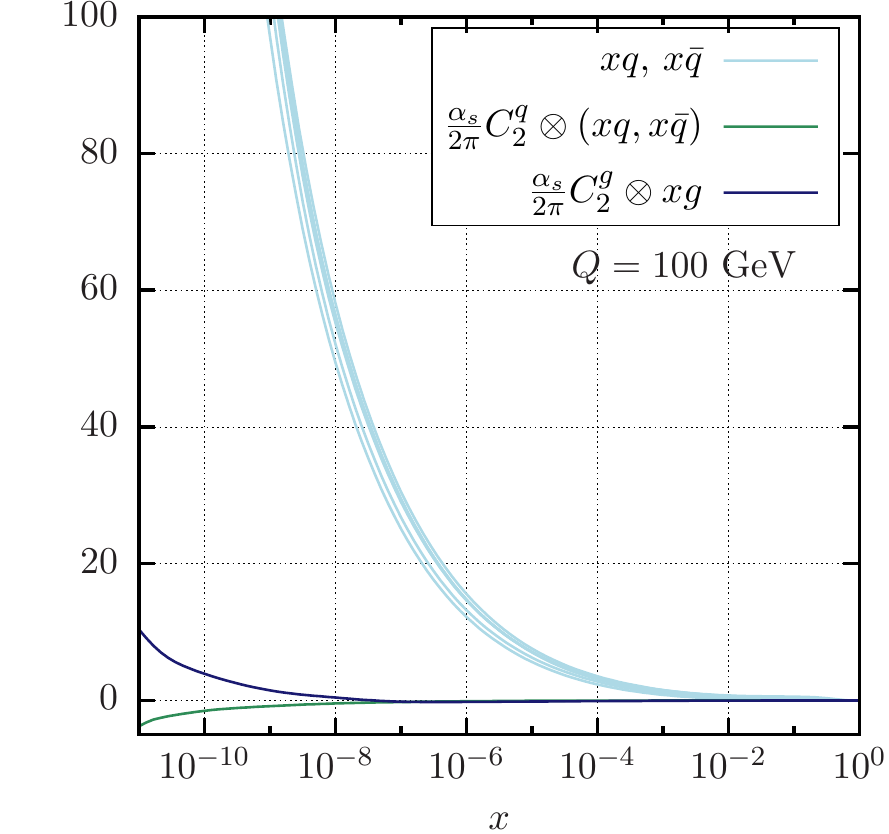}
\end{center}
\vspace*{-0.5cm}
\caption[a]{Left: Comparison of $\sigma^{\rm{LO}}\otimes \rm{PDF}^{\rm{NLO}}$ and $\sigma^{\rm{NLO}}\otimes \rm{PDF}^{\rm{NLO}}$ together with their ratio. Right: PDFs and $\dis\frac{\alpha_s}{2\pi}C_2^{f}\otimes\rm{PDF}$ of quarks and gluons as a function of $x$.}
\label{cfpdf}
\end{figure}

We have found, in agreement with \cite{Chen:2013dza,Gandhi:1998ri,Arguelles:2015wba,Dicus:2001kb,Gluck:1998js,Gandhi:1995tf}, that the purely NLO, $\alpha_s$-dependent terms in the calculation of the structure functions are negligible ($\lesssim 5 \%$) compared to the LO terms, provided that the corresponding linear combinations of PDFs appearing in the leading order expression of the structure functions, Eqs.~(\ref{cc}--\ref{nc3}), are evaluated at NLO accuracy. We have computed $\sigma_{\nu N}$ in two different ways: using the leading order expressions for the $F_i(x,Q^2)$ of Eqs.~(\ref{cc}--\ref{nc3}) with the PDFs evaluated at NLO and with the NLO expressions for $F_i(x,Q^2)$ of Eqs.~(\ref{ccnlo}--\ref{nc3nlo}). The former is labeled as $\sigma^{\rm{LO}}\otimes \rm{PDF}^{\rm{NLO}}$ in Fig.~\ref{cfpdf} and is the one commonly used in the literature, where $\alpha_s$-dependent terms are disregarded. The latter is tagged as $\sigma^{\rm{NLO}}\otimes \rm{PDF}^{\rm{NLO}}$ and corresponds to the strict calculation of the NLO cross section. The result of these two calculations are shown in Fig.~\ref{cfpdf}, left-top plot, whereas their ratio 
\eq{
R = \dis \frac{\sigma^{\rm{NLO}}\otimes\rm{PDF}^{\rm {NLO}}}{\sigma{\rm^{LO}}\otimes\rm{PDF}^{\rm{NLO}}}
}
is shown in the bottom plot. The value of $R$ stays close to unity and almost constant for all the neutrino energy range studied here. This result is better understood by exploring the relative contribution to the differential neutrino-nucleon cross section of each individual parton flavor both at LO and NLO accuracies, shown in the right plot of Fig.~\ref{cfpdf} for $Q^2=10^4$~GeV$^2$, which is the typical $Q^2$-value that dominates in this process. Due to the very strong rise of the gluon distribution function at small-$x$  -- observed in HERA data and accounted for in the PDF sets used here --, it may be expected that this contribution, which only appears in the calculation of the cross section at strict NLO accuracy, would become the dominant one at sufficiently high $E_{\nu}$. However, this is not the case as the smallness of the coupling and the convolution with the coefficient functions compensate for the growth in the gluon distribution at very small values of $x$. We conclude that the leading terms of the NLO expressions Eqs.~(\ref{ccnlo}-\ref{nc3nlo}) provide an accurate prediction of the  UHE $\nu N$ cross section.

\begin{figure}[htb]
\begin{center}
\includegraphics{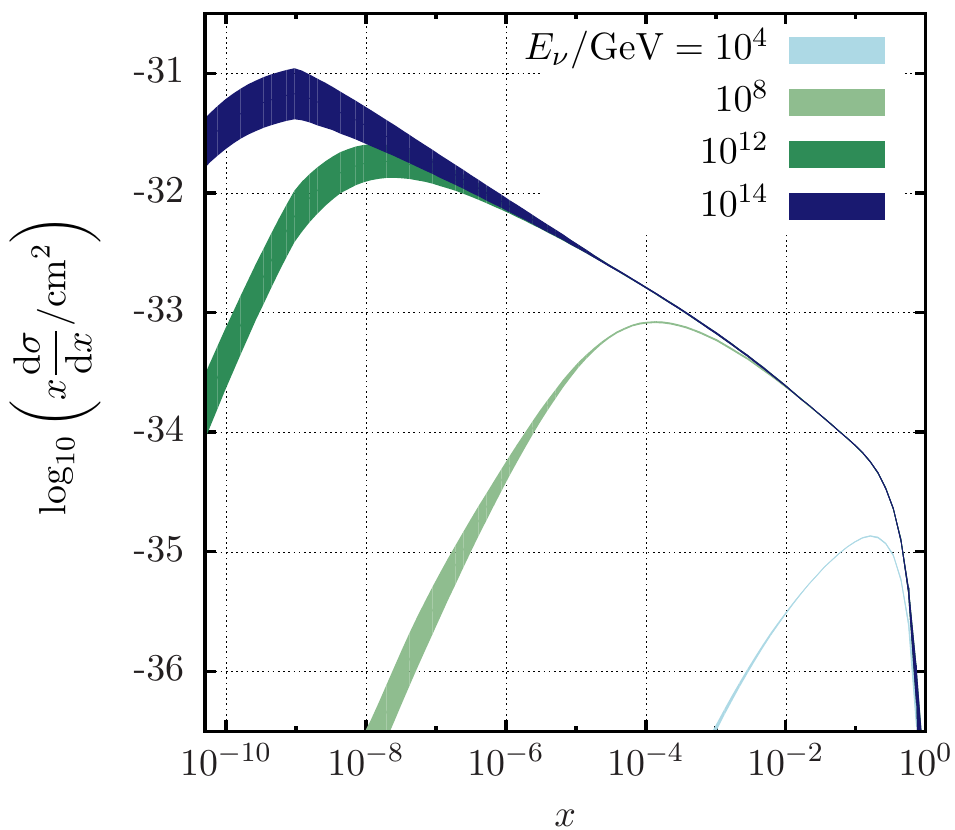}
\end{center}
\vspace*{-0.5cm}
\caption[a]{The neutrino-nucleon cross section $\dd\sigma/\dd x$ with 68~\%~C.L. error bands for $E_{\nu}=10^{4,8,12,14}$~GeV as a function of $x$ calculated with the {\tt NNPDF3.0} PDF set.}
\label{dsigmavsx}
\end{figure}

In our previous discussion about the kinematical region explored in the neutrino-nucleon interaction we have indicated that high $E_{\nu}$ is equivalent to small values of $x$. The most recent update of the experimental data available of DIS \cite{Ball:2014uwa} sets the $x$-limit in $\sim10^{-5}-10^{-6}$. The PDFs are essentially unconstrained below this barrier and these uncertainties are fully propagated to the neutrino-nucleon cross section. Smaller values of Bjorken-$x$ are accessible in the very forward region of the LHC by the LHCf or TOTEM experimental programs, but those data are associated to very small values of the virtuality $Q^2\lesssim 1$~GeV $^2$ and, hence, fall outside the domain of applicability of the DGLAP approach, valid only for perturbatively large $Q^2$-values.
In Fig.~\ref{dsigmavsx} we can observe how the $x$-value providing the largest contribution to the total neutrino-nucleon cross section shifts towards smaller values when increasing the neutrino energy. The uncertainty in the PDFs in that region is directly translated into the error bands for the cross section, given at 68\%~C.L. As an example, for the highest neutrino energy we find that the dominant $x$-value is $\sim 10^{-9}$ and the PDFs have a $\sim 20 \%$ error.  

\begin{figure}[htb]
\begin{center}
\includegraphics[scale=0.85]{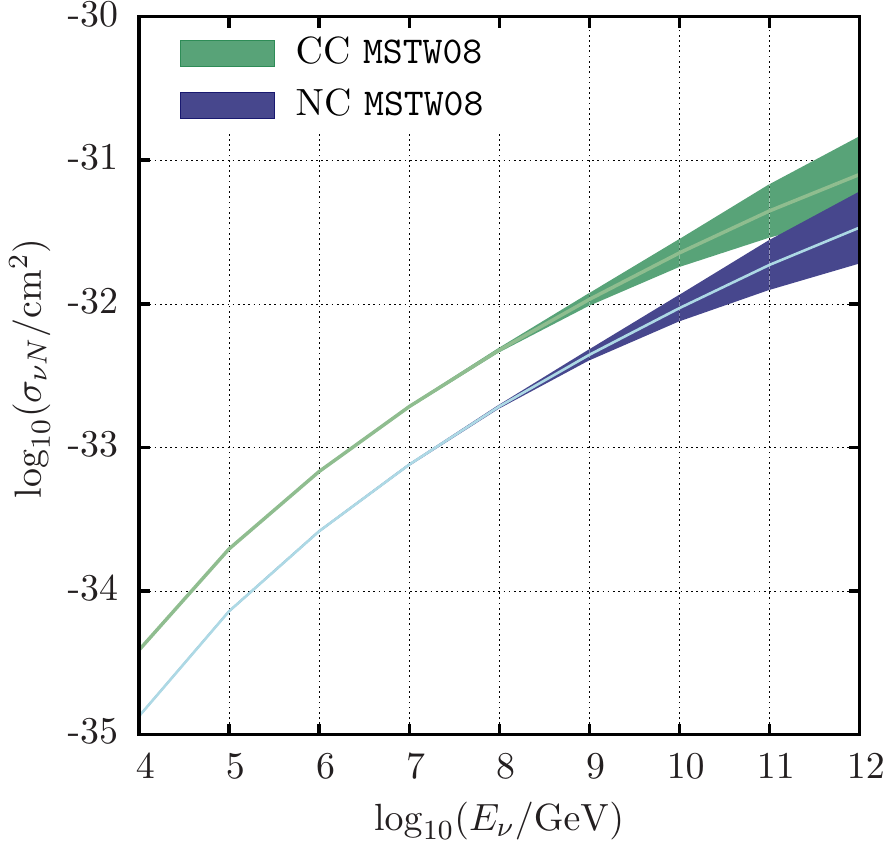}
\includegraphics[scale=0.85]{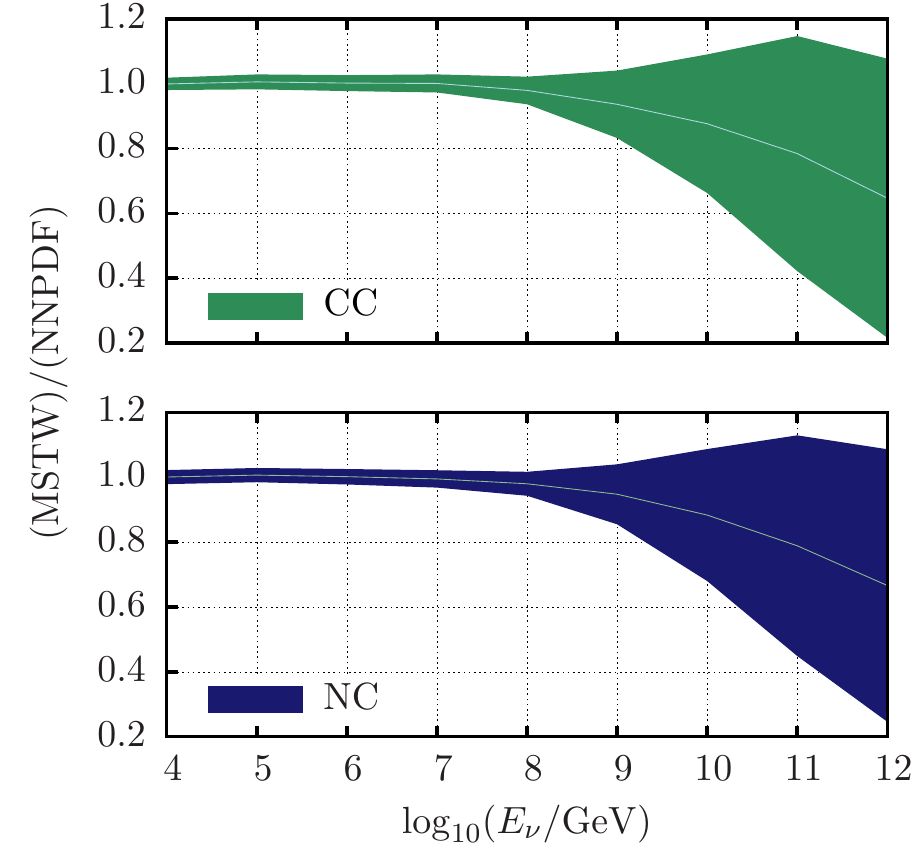}
\end{center}
\vspace*{-0.5cm}
\caption[a]{Left: the neutrino-nucleon cross section at NLO with {\tt MSTW08} PDFs. Right: comparison between the results for $\sigma_{\nu N}$ obtained with {\tt MSTW08} and {\tt NNPDF3.0}.}
\label{mstwvsnnpdf}
\end{figure}

This uncertainty, however, is not universal for all the PDFs collaborations even though they use an analogous procedure: an initial parametrization for the PDF that is fitted to data followed by DGLAP evolution. Not only the error bands but also the central value changes from one set to another, as is shown in Fig.~\ref{mstwvsnnpdf}. We show the neutrino-nucleon cross section calculated with the  {\tt MSTW08} PDF set in the left plot of Fig.~\ref{mstwvsnnpdf}. It can be seen how the error bands increase while rising the neutrino energy due to the increasing contribution of small-$x$ values reaching a $30 \%$ uncertainty for the highest energy. The right plot shows the ratio between the cross section obtained with the {\tt MSTW08} and {\tt NNPDF3.0} PDF sets, both for charged and neutral currents. The prediction stemming from the {\tt NNPDF3.0} set is higher for $E_{\nu}\gtrsim 10^8$~GeV becoming incompatible with the {\tt MSTW08} set within the error band at the highest neutrino's energy. This illustrates the decreasing predictive power of the approaches based on DGLAP when extended in the $x$-direction to values smaller than those included in their fitting data sets. Moreover, the {\tt MSTW08} PDF set covers values down to {\it only} $x=10^{-9}$, which sets an upper value of $E_{\nu}=10^{12}$~GeV for the energies that can be studied within this parametrization. In turn, the NNPDF routines provide access to $x$-values  down to $x=10^{-11}$ and beyond, allowing to extend the energy range of our study up to $E_{\nu}=10^{14}$~GeV. Furthermore, the error bands of {\tt NNPDF3.0} are systematically smaller than the {\tt MSTW08} ones. Thus, in what follows all the DGLAP based results will be calculated using the {\tt NNPDF3.0} PDF set. 

\subsection{Dipole model and BK evolution}\label{bk}

In the dipole model of DIS at low-$x$, the neutrino-nucleon cross section affords the following physics interpretation: long before reaching the target, the exchanged electroweak boson fluctuates into a colourless quark-antiquark dipole. Subsequently, the $q\bar{q}$ dipole scatters off the hadronic target via multiple gluon exchanges. The DIS structure functions for neutrino-nucleon scattering in the dipole model are given by
\eq{
F_{\rm{T,L}}^{\rm{CC,NC}}(x,Q^2) = \dis\frac{Q^2}{4\pi}\int \dd^2 {\bf{b}} \dd^2 {\bf{r}}\int_{0}^{1}\dd z \left|\psi_{\rm{T,L}}^{W^{\pm},Z^{0}}(z,Q^2,{\bf{r}})\right|^2 \mathcal{N}_{\rm{dip}}(x,\bf{r},\bf{b})
\label{dip}
} 
where $z$ is the fraction of longitudinal momentum of the gauge boson carried by the quark, $\bf{r}$ is the transverse separation between the quark and the antiquark and $\bf{b}$ the impact parameter of the dipole-target collision (henceforth boldface notation indicates two-dimensional vectors). The expressions of the wave functions $\psi_{\rm{T,L}}$ for the splitting of the gauge boson, with transverse (T) or longitudinal (L) polarizations, into a $q\overline{q}$ dipole at lowest order are given in Appendix \ref{dipole}. All the information about the strong interactions in Eq.~(\ref{dip}) is encoded in $\mathcal{N}_{\rm{dip}}(x,\bf{r},\bf{b})$, the dipole-hadron scattering matrix. Under the translational invariant approximation, i.e. assuming that the nucleon is homogenous in the transverse plane, it only depends on the absolute value of the dipole transverse size, $r\equiv |\bf{r}|$ such that the dipole-hadron cross section is  given by
\eq{
\sigma_{q\bar{q}}(x,r)=2\int \dd^2 {\bf{b}}\,  \mathcal{N}_{\rm{dip}}(x,{\bf{r}},{\bf{b}}) = 2\sigma_0\,  \mathcal{N}_{\rm{dip}}(x,r)\ ,
}
where $\sigma_0$ is a parameter that fixes the normalisation and is fitted to experimental data.
Although of ultimate non-perturbative origin, the evolution of the dipole scattering amplitude towards smaller values of $x$ can be studied perturbatively via the BK equation. It reads 
\eq{
\frac{\partial\mathcal{N}(r,x)}{\partial\ln(x_{0}/x)}\!=\!\int \dd^{2} {\bf r}_{1}\mathcal{K}(r,r_{1},r_{2})\left[\mathcal{N}(r_{1},x)+\mathcal{N}(r_{2},x)-\mathcal{N}(r,x)-\mathcal{N}(r_{1},x)\mathcal{N}(r_{2},x)\right]\,
\label{bkeq}
}
where $x_0$ is the initial value of Bjorken-$x$ and $\mathcal{K}$ is the evolution kernel, which plays an analogous role to the DGLAP splitting functions. The evolution kernel is now known to NLO accuracy in $\alpha_s \ln1/x$~\cite{Balitsky:2008zza} and also at running coupling accuracy \cite{Balitsky:2006wa,Kovchegov:2006vj,Albacete:2007yr}. However, NLO BK evolution turns unstable for a large class of initial conditions~\cite{Lappi:2015fma}. Therefore, here we shall solve the BK equation either at running coupling accuracy and also adding double logarithmic corrections recently calculated in \cite{Beuf:2014uia,Iancu:2015vea} to the evolution kernel. These two evolution schemes proceed through an all-order resummation of just a subset of terms appearing at NLO accuracy, thus rearranging the perturbative series. 

\begin{figure}
\begin{center}
\includegraphics{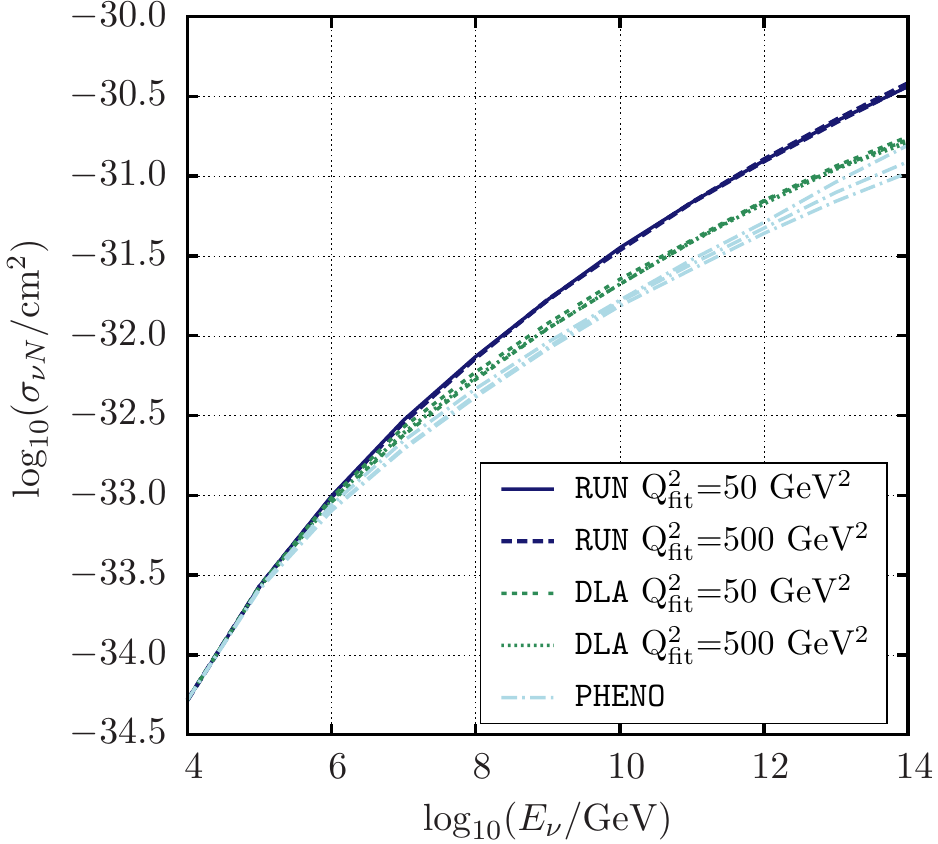}
\end{center}
\vspace*{-0.5cm}
\caption[a]{The neutrino-nucleon cross section calculated within the dipole model using different parametrizations  for the dipole cross section, $\sigma_{q\bar{q}}$: {\tt RUN}, {\tt DLA} and {\tt PHENO}. }
\label{dip-xs}
\end{figure}

\begin{figure}
\begin{center}
\includegraphics{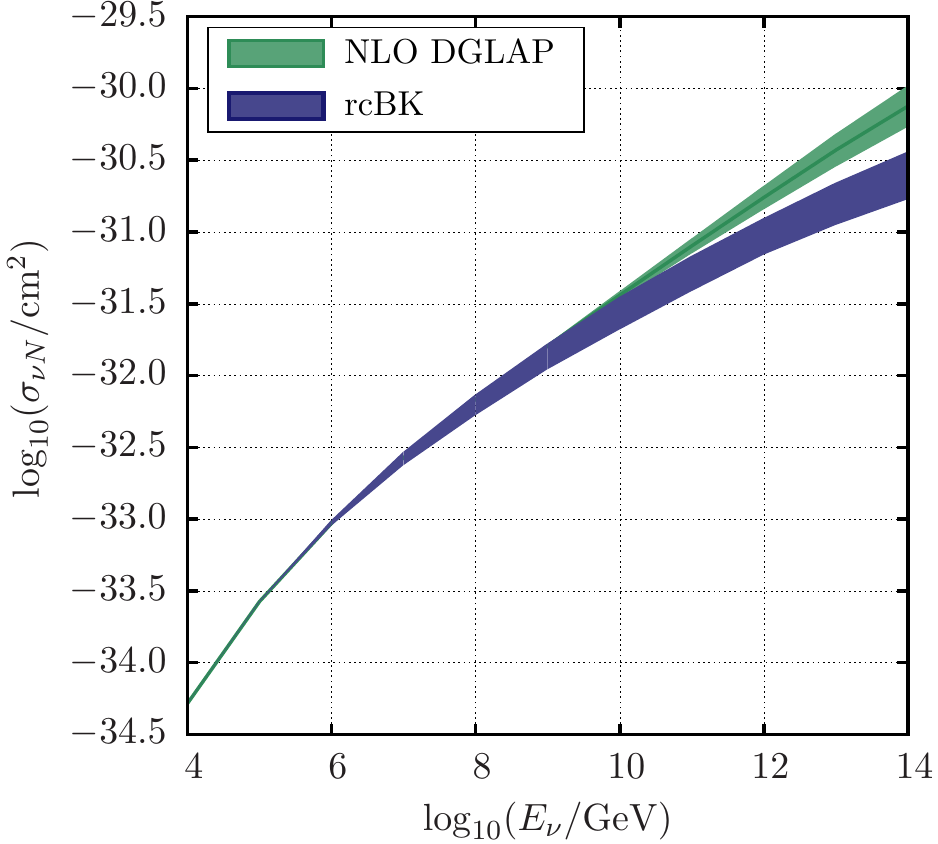}
\end{center}
\vspace*{-0.5cm}
\caption[a]{Neutrino-nucleon cross section calculated with NLO DGLAP with {\tt{NNPDF3.0}} and with running coupling BK.}
\label{final}
\end{figure}

Similar to DGLAP evolution, solving the BK equation is an initial value problem, i.e. it is well defined only after initial conditions at the initial evolution scale $x_0$ have been provided for all values of the dipole size $r$. This introduces free parameters to be fitted to data. Here we shall use the parametrizations of the dipole scattering data obtained in References \cite{Albacete:2009fh,Albacete:2010sy,Albacete} after fitting all available data on the reduced cross sections measured in $ep$ collisions at HERA with $x<10^{-2}$. In these works the evolution kernel was evaluated at running coupling accuracy \cite{Albacete:2009fh,Albacete:2010sy}, yielding a very good description of the data $\chi^2/\text{d.o.f}\sim 1$. We shall refer to this parametrization as {\tt RUN} in what follows. It should be noted that a study of the neutrino-nucleon cross section using running coupling BK evolution has been presented in \cite{Goncalves:2010ay}. More recently a new parametrization of the dipole amplitude tested against the same data set but also including large double transverse logarithmic corrections in the evolution kernel \cite{Iancu:2015vea} (also referred to as kinematic corrections in \cite{Beuf:2014uia}) has been presented in \cite{Albacete}. Again, the reported fit results were very good with $\chi^2/\text{d.o.f}\sim 1$ in all cases. We shall refer to this parametrization as {\tt DLA} in what follows. The main effect of including double logarithmic corrections in the evolution kernel is a further slowdown of the evolution speed towards small-$x$. As a consequence the {\tt DLA} parametrization yields a systematically smaller value of the neutrino-nucleon cross section than the {\tt RUN} parametrization.  

The maximum values of virtuality included in both the {\tt RUN} and {\tt DLA} fits is $Q^2=500$~GeV$^2$. Therefore the initial parametrization are not constrained in the region  $Q^2=10^4$~GeV$^2$ relevant for our calculations. This is an analogous problem to the extension of DGLAP PDF fits to the region of very small-$x$ previously discussed. An important difference, though, is that at high rapidities $Y\!=\!\ln (x_0/x)$, and on account of its non-linear character, the solutions of the  BK equation become independent of the initial conditions. In other words, the existence of a fixed point in BK evolution reduces the dependence on the fit parameters, resulting in asymptotically universal predictions that are only controlled by the dynamical information contained in the evolution kernel. A detailed study of this feature of BK evolution, referred to as {\it {scaling}} in the literature  can be found in \cite{Albacete:2004gw, Kuokkanen:2011je}.  We have tested the sensitivity to the initial conditions by reducing the maximum $Q^2$ included in the HERA data fitting set from 500 to 50 GeV$^2$. This implies a change in the initial conditions for the evolution. We observe that, for a given evolution scheme (running coupling only or running coupling plus DLA corrections), this results in negligible difference in the calculation of the $\nu N$ cross sections.

The evaluation of the total neutrino nucleon cross section involves integration of the differential cross section (Eq.~(\ref{dif-xs})) over all values of Bjorken-$x$, $0<x<1$, whereas the dipole amplitude parametrization used here are only well defined for $x<10^{-2}$. Other works in the literature circumvent this problem by extending analytically the dipole amplitudes into the region $x>10^{-2}$. Here, in order to account for the contributions to the cross sections from the region $10^{-2}<x<1$ we will assume NLO DGLAP evolution for this range of $x$. Furthermore, we shall fix the normalisation $\sigma_0$  imposing that the BK-cross section coincides with the DGLAP one at the lowest neutrino energy we are considering, $E_{\nu}=10^{4}$~GeV. This choice is well motivated since the $(x,Q^2)$-region proben at that energy is well described by the DGLAP formalism, that is, large $Q^2$ and moderate $x$. This procedure implies increasing the parameter $\sigma_0$ of the {\tt{RUN}} and {\tt{DLA}} parametrizations  by a $\sim 20 \%$, from $\sim 25$ mb to $\sim 30$ mb approximately.

For comparison, we shall also calculate the neutrino-nucleon cross section using the GBW~\cite{Golec-Biernat:1998js}, IIM~\cite{Iancu:2003ge} and Soyez~\cite{Soyez:2007kg} phenomenological models for the quark-antiquark dipole cross section, recently used in \cite{Arguelles:2015wba} for the calculation of the $\nu N$ cross section at UHE. Other phenomenological studies of the neutrino-nucleon cross section in different QCD approaches, including the dipole model, can be found in ref.~\cite{Armesto:2007tg}. We shall refer to these models collectively as {\tt{PHENO}} in what follows. Contrary to the  {\tt {RUN}} and {\tt {DLA}}  parametrizations,  these models were fitted to the old HERA data on structure functions, which have been superseded by the much more precise combined analyses performed by the H1 and ZEUS collaborations~\cite{:2009wt}. It is also known that more exclusive features of the {\tt PHENO} models are incompatible with  more exclusive data like, e.g. single inclusive hadron production cross sections in $pp$ collisions (see~\cite{Albacete:2014fwa} for an extended discussion). Finally, the energy dependence of  {\tt {PHENO}} models is mostly driven by a fit parameter, $\lambda$, that provides the $x$-dependence of the saturation scale $Q_s^2(x)\sim x^{-\lambda}$. Thus, their extrapolation beyond the region of phase space covered by the fitted HERA data is not supported by a well defined QCD dynamical input.

We present our results for the $\nu N$ total cross section as a function of the neutrino energy calculated within the dipole model in Fig.~\ref{dip-xs}. We observe that the {\tt {DLA}} parametrizations yield a systematically smaller cross section than then {\tt {RUN}} ones. The differences between the two are of a factor $\lesssim 2$ for $E_{\nu}=10^{14}$~GeV. This is an expected result, since the main role of DLA corrections is to further reduce the phase-space for small-$x$ gluon emission, thus resulting in a slower growth of  the dipole amplitude with decreasing values of $x$. Once the  evolution kernel for BK is fixed, the value of $\sigma_{\nu N}$ is almost insensitive to the initial conditions for the BK equations. This is seen by comparing the predictions for either {\tt{DLA}} or {\tt{RUN}} parametrizations fitted to different subsets of HERA data, up to $Q^2_{\rm{fit}}=$ 50 and 500 GeV$^2$ respectively. Finally, the results obtained from the {\tt PHENO} parametrizations yield an even smaller value of the cross section at high energies. We conclude that the main systematic uncertainty in the calculation of $\sigma_{\nu N}$ within the dipole model stems from the precise dynamical input embodied in the evolution kernel, and not from the initial conditions for the evolution. 

We summarise the results of this section in Fig.~\ref{final}. There we compare the values for $\sigma_{\nu N}$ obtained under the DGLAP and BK QCD evolution schemes. The band for the DGLAP based calculation originates from the uncertainties of the {\tt NNPDF3.0} PDF set used here. In turn, the uncertainty band associated to the BK-based results is related to the choice of running coupling or running coupling plus DLA corrections evolution schemes for the high-energy extrapolation. We see that the DGLAP and BK approaches yield similar results for $\sigma_{\nu N}$ up to energies $E_{\nu}\sim10^{8}$~GeV. For higher neutrino energies, the BK-based approaches lead to a systematically smaller value of the cross section than the DGLAP based calculation. That difference between DGLAP and BK approaches become as high as a factor $\sim 4.5 $ at the highest neutrino energy $E_{\nu}=10^{14}$~GeV.

\section{Limits on astrophysical neutrino fluxes}

In absence of a unified theoretical framework for the description of QCD scattering in the full kinematic $(x,Q^2)$-plane, the differences in the theoretical prediction of the total neutrino-nucleon cross section induce an uncertainty in the determination of the upper bounds of astrophysical neutrino fluxes. 

The number of neutrino-induced events, in the form of penetrating air showers, at an array of water-Cherenkov surface detectors, like those of the Pierre Auger Observatory, is
\eq{
N_{\rm evt} = \int\dd E_\nu\frac{\dd\phi_\nu}{\dd E_\nu}{\cal E}_{\rm tot}(E_\nu)
\label{flux}
}
where the {\em total} exposure is calculated given the different flavor weigths $\omega_i$ in the flux, and the $\nu$-nucleon cross sections from
\eq{
{\cal E}_{\rm tot}(E_\nu) = \sum_{i=1,2,3}\sum_{\nu_i,\bar\nu_i}
\omega_{\nu_i}(E_\nu)
\int_0^1\dd y\, {\cal E}(y E_\nu)\frac{\dd\sigma_{\nu_i N}}{\dd y}\ .
\label{flux2}
}
The {\em exposures} are the product of the effective {\em aperture}, that is the  detectors projected area weighted by the detection probability integrated over solid angle, and the range of depths within which the shower must originate to trigger the device, integrated over time. They are functions of the shower energy $E_{\rm sh}=y E_\nu$ and depend on the detection method and the shower type (hadronic or electromagnetic, according to the nature of their first interaction). Importantly, they can also depend on the neutrino-nucleon cross section. In a recent work \cite{Aab:2015kma}, a detailed calculation of Auger's total exposure has been presented from the combination of three channels: Earth-skimming neutrinos (ES) and down-going neutrinos from high angles (DGH) and low angles (DGL). The first one involves $\nu_\tau$ CC interactions, and the other two all flavors with both CC and NC interactions. Assuming the same flux of all flavors, $\phi_\nu/2\equiv\phi_{\nu_i}=\phi_{\bar\nu_i}$, with
\eq{
\frac{\dd\phi_\nu}{\dd E_\nu} = k\,E_\nu^{-2}
}
and an upper limit of $N_{\rm up}=2.39$ signal events at 90\% C.L., from zero observed and zero background events, including the uncertainties in the exposures, they set a limit
\eq{
k_{90}<\frac{N_{\rm up}}{\int\dd E_\nu\,E_\nu^{-2}{\cal E}_{\rm tot}(E_\nu)}
=6.4\times10^{-9}\ {\rm GeV}\,{\rm cm}^{-2}\,{\rm sr}^{-1}\,{\rm s}^{-1}\ .
} 
This limit applies in the energy interval $10^8<E_\nu/{\rm GeV}<2.5\times10^{10}$ and was obtained adopting the DGLAP $\nu$-nucleon cross sections given in \cite{CooperSarkar:2007cv}. The relative contributions of the three channels to the expected rate are (ES):(DGH):(DGL)~$\sim$ 0.84:0.14:0.02. According to our running coupling BK calculation with DLA corrections (see Fig.~\ref{final}) their result may be based on an overestimated expectation for the event rate. Neglecting in this case the dependence of the exposures on the neutrino-nucleon cross-section in Eqs. \ref{flux} and \ref{flux2}, we obtain that the limit $k_{90}$ could be enlarged by a factor of $\sim 1.5$, that is the inverse of the ratio between the BK cross section and that of DGLAP at an intermediate value of $E_{\nu}=10^9$~GeV.

Over an energy interval, $1.6\times10^6<E_\nu/{\rm GeV}<3.5\times10^9$, IceCube \cite{Aartsen:2013dsm} has established an upper bound on the same flux,
\eq{
k_{90}<8.3\times10^{-9}\ {\rm GeV}\,{\rm cm}^{-2}\,{\rm sr}^{-1}\,{\rm s}^{-1}
}
that should be rescaled similarly.

Even higher energies is the domain of neutrino radio-detection experiments, based on the coherent radiation emitted by secondary charged particles travelling faster than the phase velocity of light in a dense dielectric, radio-transparent medium, such as ice, sand or the lunar regolith (Askaryan effect). A few experiments operate with terrestrial ice as target, monitored with antennas embedded in ice (RICE \cite{Kravchenko:2011im}), suspended from balloons (ANITA-2 \cite{Gorham:2010kv}) or mounted on a satellite (FORTE \cite{Lehtinen:2003xv}). They are sensitive to increasing energy intervals: $10^8<E_\nu/{\rm GeV}<10^{11}$, $10^9<E_\nu/{\rm GeV}<10^{14.5}$ and $10^{13}<E_\nu/{\rm GeV}<10^{17}$, respectively. In particular, ANITA-2 sets a 90\% C.L. limit to a pure $E_\nu^{-2}$ in the energy interval $10^9<E_\nu/{\rm GeV}<10^{14.5}$ of 
\eq{
k_{90}<2\times10^{-7}\ {\rm GeV}\,{\rm cm}^{-2}\,{\rm sr}^{-1}\,{\rm s}^{-1}
}
using the DGLAP prediction for the $\nu N$ cross sections given in \cite{Gandhi:1995tf}. Although the energy explored by ANITA-2 is higher than that in Auger, we obtain a similar enhancement factor of $\sim 1.4$ at $10^{11}$~GeV since the event rate scales with $\sigma_{\nu N}^{0.45}$ according to \cite{Gorham:2010kv}.  

There are also several experiments detecting radio pulses from neutrino-initiated particle cascades in the Moon, under the LUNASKA project \cite{James:2009sf}. They offer the largest potential aperture but they are sensitive only to the most energetic neutrinos ($E_\nu > 10^{12}$~GeV), so that the pulse is bright enough to be visible from such a long distance. Based on a geometric aperture $A(E)$, a 90\% C.L. model-independent limit to a diffuse isotropic flux,
\eq{
\frac{\dd\phi_\nu}{\dd E_\nu}<\frac{2.3}{E_\nu}\frac{1}{t_{\rm obs}A_{\rm eff}(E_\nu)}
} 
has been presented in \cite{Bray:2015lda}, where $t_{\rm obs}$ is the effective observing time. An important source of uncertainty is the cross section that translates almost linearly to the aperture and hence to the flux limit \cite{James:2009sf}. The flux limits from these experiments can be up to a factor of 2.5--4.5 larger in the energy range from $E_{\nu}=10^{12}$--$10^{14}$~GeV, following BK-based result for $\sigma_{\nu N}$ presented in this work.  

\section{Conclusions}

We have presented a quantitative study of the neutrino-nucleon total cross section in two different QCD approaches: NLO DGLAP and running coupling BK evolution. We have used state-of-the art parametrizations of the parton distribution functions, in the DGLAP case, and of the dipole-nucleon scattering amplitude in the BK case. These parametrizations have been successfully tested against the available experimental data from HERA and the LHC. The differences in the cross sections arising from the use of one or another approach are due to the different theoretical input driving the extrapolation to the previously unexplored kinematic territory of UHE neutrino-nucleon collisions. The cross sections obtained in the BK framework are systematically smaller than those calculated within the DGLAP framework, the difference between these two approaches being as large as a factor 4.5 at the highest neutrino energies studied in this work $E_{\nu}=10^{14}$~GeV. This is an expected result, since the BK framework includes dynamical recombination effects that reduce the growth of the gluon densities, and hence of the total cross section as well, at small values of Bjorken-$x$. These systematic differences in the theoretical calculation of the total neutrino-nucleon cross section affect directly the experimental analyses for the determination of the neutrino fluxes of astrophysical origin. Neutrino experiments exploring the extremely high energy range are the best playground to test our predictions, that start to differ significantly from those of DGLAP for $E_\nu \gtrsim 10^8$~GeV. We have shown that for the energies explored by the LUNASKA experiments, $E_\nu \sim 10^{14}$~GeV, the actual predictions for the upper bounds could be enlarged by a factor of up to $\sim 4.5$.

\section*{Acknowledgements}
We would like to thank Antonio Bueno, Manuel Masip, Sergio Navas and Gonzalo Parente for many helpful and illuminating discussions. This work has been supported by the Spanish MINECO (FPA2013-47836 and Consolider-Ingenio MultiDark CSD2009-00064) and by Junta de Andaluc{\'\i}a (FQM101, 3048, 6552). The research of J.L.A. is also funded by a Ram\'on y Cajal contract of MINECO and by the {\it QCDense} Career Integration Grant of the FP7 Framework Program for Research of the European Commission, reference number CIG/631558. 

\appendix
\section{LO and NLO DGLAP expressions for $\nu N$ scattering}\label{nlo}

\renewcommand{\baselinestretch}{1.4}
\begin{table}
\centering
\begin{tabular}{|c||c|c|}
\hline
$f$ & $g^f_L=T^{f_L}_3-Q_f\sin^2\theta_W$ & $g^f_R=-Q_f\sin^2\theta_W$ \\
\hline\hline
$u$ & $\frac{1}{2}-\frac{2}{3}\sin^2\theta_W$ & $-\frac{2}{3}\sin^2\theta_W$ \\
$d$ & $-\frac{1}{2}+\frac{1}{3}\sin^2\theta_W$ & $\frac{1}{3}\sin^2\theta_W$ \\
\hline
\end{tabular}
\caption{Relevant couplings for neutral current interactions with $\theta_{W}$ the weak mixing angle.\label{couplings}}
\end{table}
\renewcommand{\baselinestretch}{1}

At leading order (LO) we find \cite{devenish2004deep} the following structure functions for charged currents,
\eq{
\label{cc}
2x F_1^{\rm{CC}}(x,Q^2) &= F_2^{\rm{CC}}(x,Q^2) \\
\label{cc2}
F_2^{\rm{CC}}(x,Q^2)&=x\left[\dis\frac{1}{2}\left(u+d+\overline{u}+\overline{d}\right)+s+c+b\right] \\
\label{cc3}
F_3^{\rm{CC}}(x,Q^2)&=\frac{1}{2}\left(u+d-\overline{u}-\overline{d}\right)+s-c+b
}
and for neutral current interactions,
\eq{
\label{nc}
2x F^{\rm{NC}}_1(x,Q^2) &= F^{\rm{NC}}_2(x,Q^2) \\
\label{nc2}
F^{\rm{NC}}_2(x,Q^2)&=
x\bigg\{\dis\frac{1}{2}\left[(g_{L}^u)^2+(g_{R}^u)^2+(g_{L}^d)^2+(g_{R}^d)^2\right]
\left(u+d+\overline{u}+\overline{d}\right) \nonumber\\
&\qquad+2\left[(g_{L}^d)^2+(g_{R}^d)^2\right]
\left(s+b\right)+2\left[(g_{L}^u)^2+(g_{R}^u)^2\right]c\bigg\}
\\
\label{nc3}F_3^{\rm{NC}}(x,Q^2)&=\frac{1}{2}\left[(g_{L}^u)^2-(g_{R}^u)^2+(g_{L}^d)^2-(g_{R}^d)^2\right]\left(u+d-\overline{u}-\overline{d}\right)
}
where the couplings are given in Table~\ref{couplings}.
To account for NLO corrections one must add a convolution of the PDFs with appropriate coefficient functions, including a contribution from gluons. 
The structure functions at NLO for $n_f$ active flavors read \cite{Basu:2002uu},
\eq{
\label{ccnlo}
F_1^{\rm{CC}}(x,Q^2)&=\left.F_1^{\rm{CC}}\right|_{\rm LO}+\dis\frac{\alpha_s(Q^2)}{2\pi}\left[C_1^q\otimes\left.F_1^{\rm{CC}}\right|_{\rm LO}+\frac{1}{2}n_fC_1^g\otimes g\right] \\
\label{cc2nlo}
F_2^{\rm{CC}}(x,Q^2)&=\left.F_2^{\rm{CC}}\right|_{\rm LO}+\dis\frac{\alpha_s(Q^2)}{2\pi}\left[C_2^q\otimes\left.F_2^{\rm{CC}}\right|_{\rm LO}+n_fC_2^g\otimes xg\right] \\
\label{cc3nlo}
F_3^{\rm{CC}}(x,Q^2)&=\left.F_3^{\rm{CC}}\right|_{\rm LO}+\dis\frac{\alpha_s(Q^2)}{2\pi}\left[C_3^q\otimes \left.F_3^{CC}\right|_{\rm LO}\right]
}
\eq{
\label{ncnlo}
F_1^{\rm{NC}}(x,Q^2)&=\left.F_1^{\rm{NC}}\right|_{\rm LO}+\dis\frac{\alpha_s(Q^2)}{2\pi}\left[C_1^q\otimes \left.F_1^{\rm{NC}}\right|_{\rm LO}+\sum_f^{n_f}[(g_L^f)^2+(g_R^f)^2]C_1^g\otimes g\right] \\
\label{nc2nlo}
F_2^{\rm{NC}}(x,Q^2)&=\left.F_2^{\rm{NC}}\right|_{\rm LO}+\dis\frac{\alpha_s(Q^2)}{2\pi}\left[C_2^q\otimes \left.F_2^{\rm{NC}}\right|_{\rm LO}+2\sum_f^{n_f}[(g_L^f)^2+(g_R^f)^2]C_2^g\otimes xg\right] \\
\label{nc3nlo}
F_3^{\rm{NC}}(x,Q^2)&=\left.F_3^{\rm{NC}}\right|_{\rm LO}+\dis\frac{\alpha_s(Q^2)}{2\pi}\left[C_3^q\otimes \left.F_3^{\rm{NC}}\right|_{\rm LO}\right]
}
with
\eq{
\sum_f^{n_f}[(g_L^f)^2+(g_R^f)^2]=\left\{2\left[(g_{L}^u)^2+(g_{R}^u)^2\right]+3\left[(g_{L}^d)^2+(g_{R}^d)^2\right]\right\}
}
for $n_f=5$, where the PDFs are evaluated at NLO. The coefficient functions are given by \cite{devenish2004deep,Altarelli:1979ub},
\eq{
C_i^{q}(z)=\dis\frac{4}{3}&\left\{2\dis\frac{\ln(1-z)}{1-z}\bigg|_{+}-\dis\frac{1+z^2}{1-z}\ln z -\dis\frac{3}{2}\dis\frac{1}{1-z}\bigg|_{+}\right. \\
&\left.+3+2z-(1+z)\ln(1-z)-\delta(1-z)\left(\dis\frac{9}{2}+\dis\frac{\pi^2}{3}\right)+\Delta _i^q\right\}
}
where
\eq{
\Delta_1^q=-2z, \hspace{3mm} \Delta_2^{q}=0, \hspace{3mm} \Delta_3^q=-(1+z)
}
and for the gluon initiated processes,
\beq
C_i^g(z)=\dis\frac{1}{2}\left\{\left[(1-z)^2+z^2\right]\ln\left(\dis\frac{1-z}{z}\right)-8z^2+8z-1+\Delta_i^g\right\}
\eeq
where
\eq{
\Delta_1^g=-4z(1-z), \hspace{3mm} \Delta_2^g=\Delta_3^g=0
}
and
\beq
\alpha_s(Q^2)=\dis\frac{\alpha_s(M_Z^2)}{1+\alpha_s(M_Z^2)\beta_0\ln(Q^2/M_Z^2)}, \hspace{5mm} \beta_0=\dis\frac{(33-2N_f)}{12\pi}\ .
\eeq
A numerically small disagreement with Ref.~\cite{Basu:2002uu} was found.
The plus prescription is defined as usual\footnote{$\dis\int_x^1\dd z f(z)g(z)_+=\dis\int_x^1\dd z[f(z)-f(1)]g(z)-f(1)\int_{0}^{x}\dd z g(z)$.} and $\otimes$ denotes the convolution defined as
\beq
C_f\otimes{\rm{PDF}}=\int_{x}^{1}\dis\frac{\dd z}{z}C_f(z){\rm{PDF}}\left(\dis\frac{x}{z}\right)\ .
\eeq

\section{Light cone wave functions for the dipole model}\label{dipole}
Explicit expressions for the wave functions squared in the massless quark limit are as follows \cite{Kutak:2003bd},
\eq{
\left|\psi^{W^{\pm}}_{\rm{T}}(r,z,Q^2)\right|^2&=\dis\frac{6}{\pi^2}[z^2+(1-z)^2]\epsilon^2K_1^2(\epsilon\bf{r}) \\
\left|\psi^{W^{\pm}}_{\rm{L}}(r,z,Q^2)\right|^2&=\dis\frac{24}{\pi^2}z^2 (1-z)^2 Q^2 K_0^2(\epsilon\bf{r}) \\
\left|\psi^{Z^{0}}_{\rm{T}}(r,z,Q^2)\right|^2&=\dis\frac{3}{2\pi^2}\left[(g_{L}^u)^2+(g_{R}^u)^2+(g_{L}^d)^2+(g_{R}^d)^2\right][z^2+(1-z)^2]\epsilon^2K_1^2(\epsilon\bf{r}) \\
\left|\psi^{Z^{0}}_{\rm{L}}(r,z,Q^2)\right|^2&=\dis\frac{24}{\pi^2}\left[(g_{L}^u)^2+(g_{R}^u)^2+(g_{L}^d)^2+(g_{R}^d)^2\right]z^2(1-z)^2 Q^2 K_0^2(\epsilon\bf{r}) 
}
where 

\beq
\epsilon^2=z(1-z)Q^2
\eeq
and $K_{0,1}(x)$ are the modified Bessel functions.

\newpage
\bibliographystyle{JHEP-2modM}

\end{document}